# Generation of Cold Argon Plasma Jet at the End of Flexible Plastic Tube

Konstantin G. Kostov[1*], Munemasa Machida[2] and Vadym Prysiazhnyi[1]

[1]Universidade Estadual Paulista – UNESP, Campus in Guaratinguetá – FEG

Guaratinguetá, SP 12516-410, Brazil

[2]Universidade Estadual de Campinas – UNICAMP, Instituto de Física – IFGW

Campinas, SP 13083-859, Brazil

*Corresponding author: kostov@feg.unesp.br

**Abstract** This brief communication reports a new method for generation of cold atmospheric pressure plasma jet at the downstream end of a flexible plastic tube. The device consists of a small chamber where dielectric barrier discharge (DBD) is ignited in Argon. The discharge is driven by a conventional low frequency AC power supply. The exit of DBD reactor is connected to a commercial flexible plastic tube (up to 4 meters long) with a thin floating Cu wire inside. Under certain conditions an Ar plasma jet can be extracted from the downstream tube end and there is no discharge inside the plastic tube. The jet obtained by this method is cold enough to be put in direct contact with human skin without electric shock and can be used for medical treatment and decontamination.

Index Terms: Atmospheric pressure plasma jet, Nonthermal plasma, Plasma propagation in plastic tube

In the last decades, cold atmospheric pressure plasma jets (APPJs) have attracted much attention due to their versatility, low-cost operation and also ability to produce reactive chemistry at room temperature [1]. These facts make the plasma jets very attractive for applications in the biomedical field. An interesting feature of APPJs is their ability to penetrate and propagate inside small holes and flexible dielectric tubes [2]. Delivery of cold plasma through flexible tube in a specific location can be very useful for endoscopic applications in medicine, such as treatment of colorectal and pancreas cancers [3-5]. Therefore, the development of appropriate plasma sources for *in vivo* treatments has been subject of intense research. So-called plasma needle reported in [3] produced cold He plasma on the tip of a thin electrode inserted into a 10-cm-long flexible catheter. The electrode was connected directly to a RF power supply and the catheter could be bent up to 30°. This device had only limited length and since the electrode was directly connected to the power supply there was risk of electric shock. Several other groups [2, 4, 5] have reported propagation of plasma into flexible plastic tubes with different length and thickness. Usually most authors flushed thin dielectric tubes with noble gases and in some cases like in ref. [5], where a special power source was used, they can deliver plasma up to few meters distance. By using a specially designed bifilar helix electrode wrapped around a several meters long, flexible, plastic tube the authors of ref. [6] were able to generate Ar plasma along entire tube and also a plasma plume was launched from the tube end. In all these works except in ref. [3] plasma was generated along the entire extension of the plastic tube, which requires higher electric power and/or gas flow.

In this work we propose a novel method for generation of an Ar plasma jet at the downstream end of a flexible, plastic tube (up to several meters long) without actually igniting discharge inside it. A distinguishable feature of the suggested method is that the plastic tube in our experiment is equipped with a thin floating Cu wire. It is inserted few mm inside the reactor, passes freely (no special support) along the entire tube extension and terminates few millimeters before the plastic tube end. In this arrangement the Cu wire has no contact with the powered electrode and the risk of electric shock is greatly reduced. A schematic drawing of the experimental setup is shown in the Fig. 1.

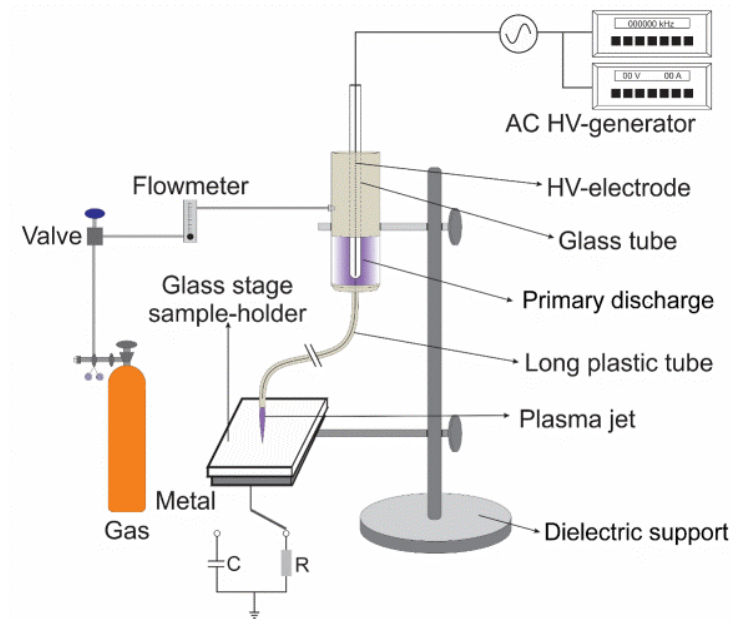

Figure 1. Schematic drawing of the experimental setup.

Besides the coaxial DBD reactor, the experimental arrangement consists of a grounded platform covered by glass, a gas admission system, a power supply and diagnostics equipment. The plasma reactor is made from a 4-cm-long glass tube (24 mm inner diameter and 2 mm thickness) whose both ends are closed. A metal rod (2.6 mm diameter) covered by a closed-end quartz tube (6.1 mm outer diameter) is fixed in the center of the glass tube and served as a powered electrode. The high voltage was generated by a commercial AC power supply Minipuls 4 (GBS Elektronik, Germany). The signal frequency was fixed at 19.0 kHz while the voltage amplitude was varied in the range of 8.0 - 23.0 kVpp. The working gas Ar (99.998%) is fed into the plasma reactor through its lateral wall and gas flow rate is controlled by a rotameter within the range 0.1-5.0 slm. The exit of the DBD reactor, located on its bottoms side, is connected to a commercial flexible plastic tube (3.2 mm outer diameter, 2 mm inner diameter), which guides the plasma species downstream. At certain conditions (especially if a conductive object is closed to the tube end) a plasma jet can be extracted from the plastic tube.

Electric characterization of the system was performed using a Tektronix TDS 3032 oscilloscope (300 MHz, 2.5 GS/s). The high voltage waveform was measured through a divider at the generator exit (1:2000). The current waveforms were obtained from the voltage drop across a grounded serial resistor of 104.7 Ohm. The voltage drop on an external capacitor of 10 nF was used for power calculation by using the Lissajous' figure method.

Depending on device operation conditions (gas flow, applied voltage, device geometry and plastic tube configuration) the plasma produced in the DBD reactor can penetrate up to ten cm inside the plastic tube and emerge from its end. Fig. 2 shows photos of Ar plasma jets formed at the end of a 10-cm-long plastic tube for two cases: (a) an empty tube and (b) a tube with a floating Cu wire inside. In both cases plasma jets are launched from the plastic tube end and spread on the glass, which covered the ground electrode. However as can be seen in the Fig. 2(a) the Ar plasma fills the entire plastic tube. The intensity of plasma-emitted radiation inside the tube without Cu wire gradually diminishes along its length and plasma jet on the tube end is weak. If the tube is made longer than 10 cm plasma jet cannot be extracted for the same applied voltage and gas flow.

It is well known that the performance of APPJs is greatly influenced by nearby conductive objects (grounded or floating) [7]. Making used of this we inserted a thin Cu wire (diameter of 0.2 mm) inside a plastic tube with the same length (10 cm). The Cu wire also has 10cm length and penetrates few mm inside the DBD chamber, and terminates few mm before the plastic tube end. The wire passes freely through the tube and no efforts were made to keep it coaxial. With floating wire inside the reactor the gas breakdown voltage is reduced and the discharge predominantly occurs between the quartz tube and the Cu tip. Fig 2(c) presents a close view of the discharge inside the DBD reactor where the wire tip can be seen. To get plasma outside the plastic tube without wire we needed to apply voltage of ~ 21 kVpp and gas flow of 1.6 slm that are much higher compared to same parameters in the case of a plastic tube with floating wire (14.0 kVpp and 0.8 slm ). Also, as can be seen in Fig. 2(b) there is no discharge inside the plastic tube, though an intense plasma jet is ignited at the Cu wire end. However the jet obtained by this method is hot and cannot be used for decontamination only for material treatment.

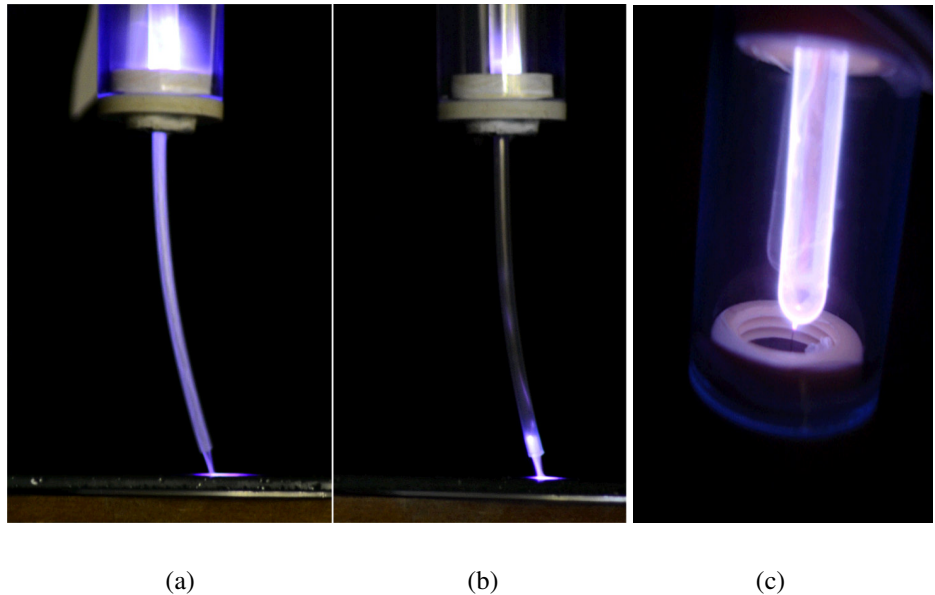

(a)          (b)          (c)

Figure 2. Comparison between plasma jets ignited in a plastic tube (a) without Cu wire, (b) with Cu wire, and (c) close view of the discharge between quartz tube and Cu wire tip. Operating conditions: (a) 21 kVpp and Ar flow of 1.6 slm and (b) and (c) 14 kVpp and Ar flow of 0.8 slm.

Furthermore, with a floating wire placed inside, it is possible to ignite plasma jet at the end of tube much longer tube (up to few meters). This is shown in Fig. 3 where a quite nice plasma jet was obtained for a tube with 160cm length. The tube in the Fig. 3(a) was bended only for better visualization of the entire system because the plasma jet operation is regardless of the tube alignment in space (straight or bended). It should be noted that there is no plasma inside the tube. However, if a metal object is placed in contact with the tube in that location plasma is ignited.

For longer tubes the jet on the downstream end becomes less intense and beyond 4 meter is not possible to ignite plasma anymore. This means the electric field at the wire tip is below the breakdown value in Ar. On the other hand for tubes longer than 1.5m the plasma jet is cold enough to be applied to human skin without thermal sensation or electric shock. This is shown in the photo in Fig. 3(b) where jet tube length of 160 cm and applied voltage of 10.0 kVp-p were used. The power of the plasma jet in this configuration was 4 W. Operating at higher voltage or shorter tube plasma jet start to become hot and also sensation of slight electric shock could be felt.

Different tube lengths and diameters as well as tube materials are possible. Also, the device can operate with other inert gases and different configurations of the "primary" plasma reactor. More details of the suggested device will be given in the following work.

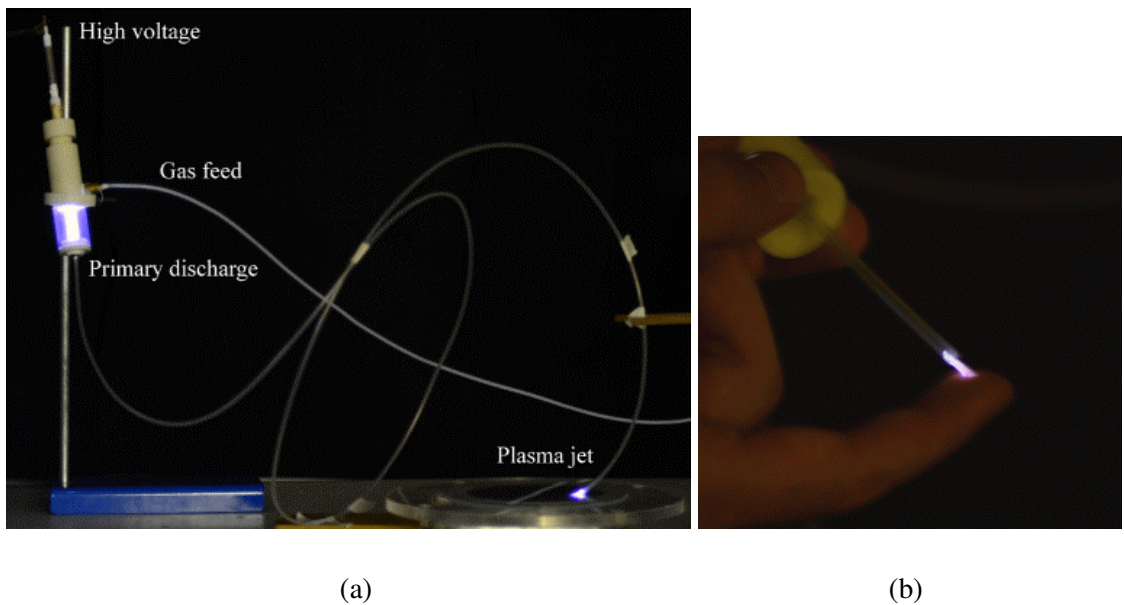

(a) (b)

Figure 3. Photos of: (a) the experimental setup and (b) the Ar plasma jet on the finger of one of the authors. Applied voltage 12 kVp-p, Ar flow 0.8 L/min, plastic tube length 160 cm.

This essentially implies some limitations to the system of transferred jet. First of all there is a maximum tube length above which plasma jet cannot be ignited. Also there are a minimum tube length and maximum voltages when the generated plasma is at about room temperature and can be put into direct contact with heat sensitive or biological objects. Moreover, in this condition the transferred jet can be applied for material modification. Nevertheless we believe still that there is a sufficiently large window of jet operation where jet can be used for decontamination device. The advantage of this jet transfer system is that it is flexible and can be easily handled.

In Summary, the possibility to remotely generate cold Ar plasma jet through a long (up to 4 meters long) flexible plastic tube has been demonstrated. The device operation relays on a thin floating Cu wire installed inside the plastic tube. Usually when a primary discharge is ignited inside the DBD reactor no plasma is formed inside the plastic tube. However under certain conditions a small Ar plasma jet is launched from the tube end. The presented setup allows separating all dangerous high voltage elements from the contact with operator, thus making plasma operation safe. The position of Cu wire inside the plastic tube allows to safely touch the plastic tube avoiding electric shock.

There is a window of operation parameters (applied voltage, tube length, gas flow rates) where it is possible to generate room temperature plasma at the end of the plastic tube, which can be used for selective treatments of biological objects.

**Acknowledgements**

Financial support from FAPESP (process 2013/06732-3) and CNPq (process 470995/2013-0) is acknowledged